# Interaction of organic surfaces with active species in the high-vacuum environment

V. Podzorov [*,1], E. Menard[2], S. Pereversev[1], B. Yakshinsky[1], T. Madey[1], J. A. Rogers[2] and M. E. Gershenson[1]

[1] Department of Physics and Astronomy, Rutgers University, Piscataway, New Jersey 08854
[2] Department of Materials Science and Engineering, University of Illinois, Urbana Champaign

(March 1st, 2005)

Using single-crystal organic field-effect transistors with the conduction channel exposed to environmental agents we have observed generation of electronic defects at the organic surface in the high-vacuum environment. Rapid decrease of the source-drain current of an operating device is observed upon exposure of the channel to the species generated by high-vacuum gauges. We attribute this effect to interaction of the organic surface with electrically neutral *free radicals* produced in the process of *hydrocarbon cracking* on hot filaments with a relatively low activation energy $E_a \sim 2.5$ eV (240 kJ/mol). The reported results might be important for optimizing the high-vacuum processes of fabrication and characterization of a wide range of organic and molecular electronic devices.

Thermal deposition of organic semiconductors in high vacuum is widely used for fabricating various organic devices, including organic thin-film transistors (OTFTs)[1] and light-emitting diodes (OLEDs)[2]. In the high-vacuum environment, the organic molecules interact with different active species generated by vacuum gauges, thermal evaporation sources, etc. These processes might lead to formation of surface electronic states (charge traps, dopants, or surface dipoles) that affect the performance of these devices. Better understanding of these phenomena is crucial for improving the performance of organic devices. Recent development of the organic field-effect transistors with the conducting channel fully accessible to environmental agents provides a unique tool to address this important issue [3,4]. Below we will refer to this type of transistors as the organic field-effect transistor with exposed channel, or OFETEC.

Using the OFETEC based on single crystals of *rubrene* and *tetracene*, we have studied the surface-restricted interactions of organic semiconductor with active species in the high-vacuum environment. It has been observed that interactions with active agents generated by high vacuum gauges and thermal deposition sources cause formation of electronic defects at organic surface and lead to deterioration of OFET's performance (the "gauge" effect). Low activation energy of generation of the active species (~ 2.5 eV) suggests that these agents might be *free radicals* created in the process of cracking of hydrocarbons on hot surfaces in vacuum systems.

Fabrication of OFETECs has been discussed in details in Ref. [3]. In these devices, the conventional solid gate dielectric is replaced by a micron-scale gap between the gate electrode and the surface of organic semiconductor (see the sketch in Fig. 1). The single crystals of rubrene and tetracene grown by physical vapor transport [5,6] were laminated against the gold source and drain contacts deposited on a patterned elastomer stamp. High quality of the single-crystal surface facilitates studies of the intrinsic mechanisms of interaction with environmental species, not complicated by surface disorder and inter-grain boundaries typical for OTFT-based sensors [7,8,9]. The conduction channel length and width are $L = 0.19$ mm and $W = 0.5 – 1$ mm; the vacuum gap is 4.5 μm. These devices have been measured over the pressure range $P = 10^{-6} – 760$ Torr in a chamber evacuated by a turbomolecular pump backed up by sorption pumps. Unless otherwise specified, all the measurements have been performed at room temperature in the dark, using Keithley K2400 source-meters and K6512 electrometers.

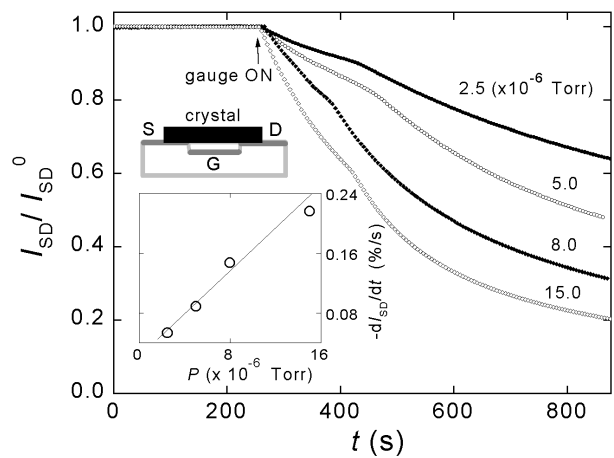

**Fig. 1.** Time evolution of $I_{SD}$ of an operating organic field-effect transistor with exposed channel - OFETEC (see the sketch) measured at different pressures $P$ in the vacuum chamber ($V_{SD} = 5$V, $V_g = -30$V). The surface of conduction channel of organic semiconductor is fully exposed to active species in the vacuum environment. A hot-cathode vacuum gauge is turned ON at $t = 250$ s. The Inset: the dependence of $I_{SD}$ decay rate (measured at $t = 300$ s) on $P$.

While testing the OFETECs in a high-vacuum chamber, we have observed that the source-drain current ($I_{SD}$), measured at fixed source-drain ($V_{SD}$) and gate ($V_g$) voltages, decreases after turning on a hot-cathode high vacuum gauge (see Fig. 1). Similar effect has been observed at Delft University[10]. This was quite unexpected, because the characteristics of our rubrene single-crystal OFETECs are very stable under high vacuum, provided that the gauge is turned off. The characteristics are also insensitive to such gases as $O_2$, $H_2O$, $N_2$, Ar, He, and air, in contrast to what is commonly observed in other organic semiconductors [8,11,12]. In addition, the bias stress effect, typical for OTFTs [13], is negligible in the OFETECs. The "gauge" effect is practically independent of the OFETEC position in the chamber and can be observed even if the device is not in a line-of-sight with the gauge. The same behavior has been observed after replacing the hot-cathode gauge with cold-cathode ones. Access of active species to the conduction channel of OFETEC was crucial for the observation of the "gauge" effect: other single-crystal OFETs, in which the surface of organic crystal is conformally coated with a parylene thin film[14], are insensitive to gauges. Although OFETEC is a powerful tool to study the surface interactions, the "gauge" effect can be also observed using 2-contact rubrene samples (without FET structure): turning on a high-vacuum gauge results in diminishing the built-in surface conductivity of rubrene crystal [15].

---

* Electronic mail: podzorov@physics.rutgers.edu

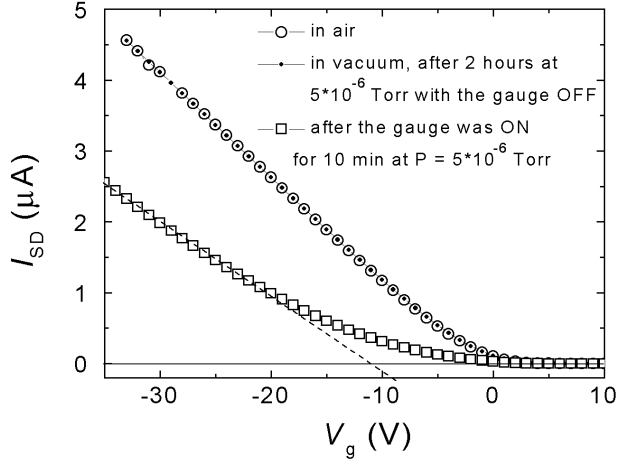

**FIG. 2.** $I_{SD}(V_g)$ characteristics of a rubrene OFETEC measured in air (circles); in vacuum after 2 hours pumping at $P = 5 \cdot 10^{-6}$ Torr with the high-vacuum gauge off (dots); and after the gauge was ON for 10 min at $5 \cdot 10^{-6}$ Torr (squares). Interaction of the conduction channel with the species generated by the gauge results in a decrease of the field-effect mobility $\mu \propto dI_{SD}/dV_g$ and increase of the threshold voltage $V_g^{th}$.

The rate of $I_{SD}$ decrease when the gauge is on depends on the pressure in the chamber $P$ (see the inset in Fig. 1): it is small at $P < 10^{-6}$ Torr, increases quasi-linearly with pressure over the range $P = 10^{-6} – 10^{-5}$ Torr, and is very high at $P \sim 10^{-4}$ Torr, where $I_{SD}$ drops by *2-3 orders of magnitude* within 1-3 sec of the gauge operation (see Fig. 4). The study of the transconductance characteristics of rubrene OFETECs (Fig. 2) shows that two effects contribute to the drop of $I_{SD}$: (a) increase of the threshold voltage $V_g^{th}$, and (b) decrease of the carrier mobility $\mu \sim dI_{SD}/dV_g$. Thus, the interaction with active species induces both shallow and deep traps at the organic surface [4]. Qualitatively similar effects are observed in tetracene single-crystal OFETECs.

We have found that the active species responsible for the "gauge" effect are *neutrals*. For this test the high-vacuum gauge has been separated from the chamber by two fine-mesh metallic grids. Biasing the grids independently with respect to the common ground with up to ±400 V did not affect the rate of the gauge effect.

Further experiments have shown that the main factor that controls the rate of $I_{SD}$ decrease is the temperature of the filament of the hot-cathode gauge, $T$, rather than the accelerating voltage between it's cathode and anode, $V_{acc}$.

The dependence of $I_{SD}$ decrease rate, $dI_{SD}/dt$, on $T$ at $V_{acc} = 0$ is shown in Fig. 3. Above $T_0 \sim 950^{\circ}$C, the rate rapidly increases and saturates at $T \sim 1200^{\circ}$C. Below $T \sim 1150^{\circ}$C, the experimental data in Fig. 3 are fitted well with an exponential dependence $dI_{SD}/dt \propto \exp(-E_a/k_BT)$ with the activation energy $E_a \sim 2.5$ eV (240 kJ/mol). We attribute this dependence to the activation process of *thermal cracking* of molecules of the residual atmosphere on the hot filament; the products of this process are *neutral free radicals* that are very reactive due to their unpaired electron.

The low activation energy suggests that the active agents are *hydrocarbon free radicals* [16]. Indeed, the measured activation energy is noticeably lower than dissociation energies of common gases present in high vacuum, such as $H_2O$, $CO$, $CO_2$ [17]. To the best of our knowledge, only the homolytic cracking of C-C or C-H bonds of heavy hydrocarbons can occur at $T$ as low as 950 $^{\circ}$C. Depending on the bond type and hydrocarbon size the homolysis energy varies over the range 150 – 420 kJ/mol [16]. We can not, however, completely rule out the possibility of other cracking processes with similar activation energies, such as atomization of molecular hydrogen ($E_a = 4.47$ eV) [17]. Saturation of $dI_{SD}/dt$ at $T > 1200$ $^{\circ}$C indicates that the probability of cracking approaches 100% at ~ 1200 $^{\circ}$C. The increase of the degradation rate with pressure (inset, Fig. 1) could be related to an increasing density of the species and their decreasing mean free path, which results in an increasing probability for a radical to get into the gap between the gate and the channel. This implies that the lifetime of the radicals is quite long at $P = 10^{-6} - 10^{-4}$ Torr.

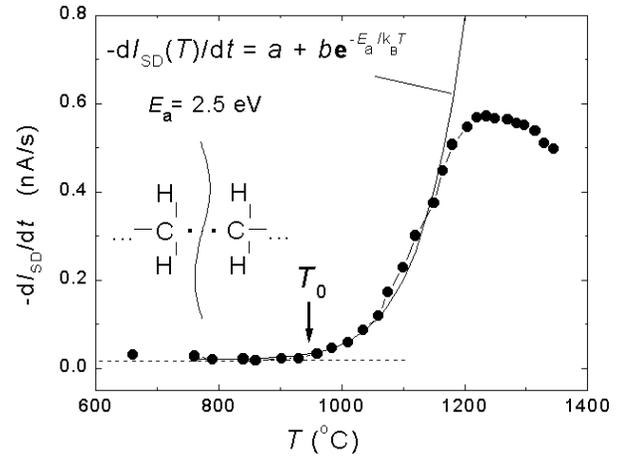

**FIG. 3.** The rate of $I_{SD}$ decrease for a rubrene OFETEC ($V_{SD} = 5V$, $V_g = -25V$) as a function of temperature $T$ of a tungsten filament resistively heated in the same vacuum chamber where the device is measured ($P = 2 \cdot 10^{-6}$ Torr). The transistor was not in a direct line of sight with the filament. The arrow indicates the onset of $I_{SD}$ decay at $T_0 \sim 950$ $^{\circ}$C. The experimental data at $T < 1150^{\circ}$C are fitted with an activation dependence (solid line) with activation energy $E_a = 2.5$ eV. The process of the homolytic thermal cracking of hydrocarbon into neutral free radicals is shown in the inset.

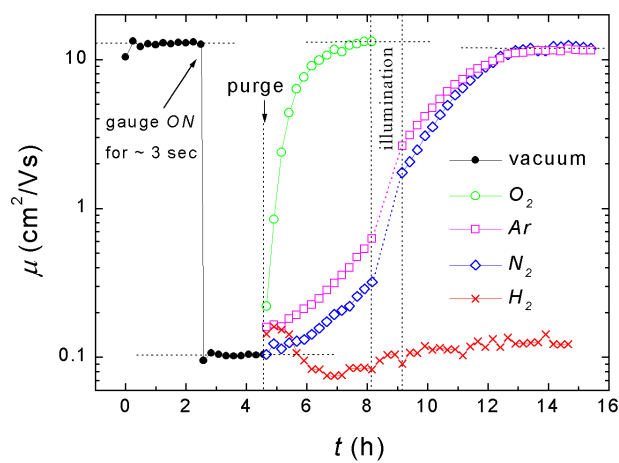

**FIG. 4.** The evolution of the mobility with time for the rubrene OFET with exposed channel measured in vacuum ($P \sim 10^{-4}$ Torr, black circles) and in 1 atm of different ultra-high-purity gases (colored symbols). The mobility has been extracted from the linear portions of $I_{SD}(V_g)$ curves measured every 15 minutes ($V_S = 5V$). A sharp drop of $\mu$ at $t = 2.5$ h was caused by turning ON a high-vacuum gauge for 3 seconds. Illumination: 100 mW/cm$^2$ of white light.

Interestingly, exposing the "corrupted" devices to the oxygen gas can restore the initially high performance of rubrene OFETECs. Figure 4 shows the mobility of a rubrene vacuum-gap OFET monitored for a period of 16 hours by measuring trans-conductance characteristics of the device every 15 min. After turning on the high-vacuum gauge for 3 seconds, the initial mobility $\mu \sim 10$ cm$^2$/V·s diminishes by 2 orders of magnitude. The device remained in this degraded state under vacuum with the gauge off. When the chamber was purged with different UHP gases, an increase of the mobility to its initial high value was observed. The fastest recovery occurs in the atmosphere of oxygen; intermediate recovery rates - in argon and nitrogen; and hydrogen causes no recovery at all. Note that all UHP gases, except $H_2$, contain some ppm concentrations of $O_2$, which is sufficient for the surface reaction to occur. As Fig. 4 shows, illumination of the device with visible light increases the recovery rate in Ar and $N_2$; qualitatively similar effect of illumination has been observed in oxygen and in air. The recovery is observed independently of the type of gauge used for the degradation, suggesting that the nature of active species created by different gauges is similar. Interestingly, no complete recovery of tetracene OFETECs has been observed.

On the basis of these observations, we can suggest the following mechanism of defect formation at the organic surface exposed to active agents in the high-vacuum environment. The free radicals generated by the gauges and chemisorbed at the organic surface create trap states with a wide energy distribution in the band gap, as evidenced by the fact that both shallow and deep traps are created. It is known that the attached radicals can be scavenged from the surface by molecular oxygen in a spin-selective reaction of radical trapping: its rate increases dramatically if the oxygen molecule is in the singlet state[18]. Therefore, the ground-state triplet $O_2$ has to be converted to the singlet $O_2$ to speed up the reaction of radical quenching. It is very likely that the observed increase of the recovery rate under illumination is due to the triplet-to-singlet conversion of $O_2$ that occurs through photosensitized energy transfer from rubrene, which is one of the most efficient organic photo-sensitizers[19]. The fact that tetracene-based devices could not be fully recovered is consistent with much poorer performance of tetracene as a photo-sensitizer.

To summarize, by using the OFETs with exposed channel as a tool for studying the surface-restricted interactions, we have observed generation of electronic defects at the organic surfaces exposed to active agents in the high-vacuum environment. We believe that these active agents are the free radicals generated in the process of hydrocarbon cracking in high vacuum gauges and on hot surfaces in vacuum chambers. The reported results are important for optimizing the high-vacuum fabrication processes of a wide range of organic and molecular electronic devices.

This work has been supported by the NSF grants DMR-0405208, ECS-0437932 and CHE 0315209. We thank E. Garfunkel and Y. Chabal for helpful discussions.